\documentstyle[11pt,newpasp,twoside,epsf]{article}
\def\edcomment#1{\iffalse\marginpar{\raggedright\sl#1\/}\else\relax\fi}
\reversemarginpar

\begin{document}
\title{High Contrast Imaging of Extrasolar Planets}
  \author{John Debes}
\affil{Penn State University University Park, PA 16802}
\author{Jian Ge}

\begin{abstract}
Gaussian aperture pupil masks (GAPMs) can in theory achieve the contrast
requisite for directly imaging an extrasolar planet.  We use lab tests and 
simulations to further study their possible place as a high contrast imaging
technique.  We present lab comparisons with traditional Lyot coronagraphs and 
simulations of GAPMs and other high contrast imaging techniques on HST.
\end{abstract}


\section{Introduction}
\label{intro}
The search to directly image an extrasolar planet requires contrast
levels of $\sim$10$^{-9}$ a few $\lambda / D$ from the central star.
Scattered light in a telescope
and the diffraction pattern of the telescope's aperture limit the contrast
possible for direct detection of faint companions.  The circular aperture of
telescopes creates a
sub-optimal diffraction pattern, the so-called Airy Pattern which is
azimuthally symmetric.  In addition, the intensity in the diffraction pattern
of the circular
aperture declines as $\theta^{-3}/\theta_o$, where $\theta_o=\lambda/D$.  
Currently the best way to diminish
the Airy pattern is to use a coronagraph by using the
combination of a stop in the focal plane that rejects a majority of the
central bright object's light and a Lyot stop in the pupil plane to reject high
frequency light (Lyot 1939; Malbet 1996; Sivaramakrishnan et al. 2001).  Several recent ideas explore the use of alternative ``apodized'' apertures for high contrast
imaging in the optical or near-infrared (Nisenson \& Papaliolios 2001; Spergel 2002; Debes, Ge, \& Chakraborty 2002).
These designs revisit concepts first experimented with in the field of optics
(Jacquinot \& Roizen-Dossier 1964).  Other designs, such as the band
limited mask, seek to null the light from a central star in much the same way 
that a nulling interfermoeter performs (Kuchner \& Traub 2002).  

By placing a mask into the pupil plane with a gaussian aperture, one can
transform a traditional circular aperture telescope into one with a diffraction
pattern better suited for high contrast imaging.  Using a mask represents a
quick, efficient, and cheap way to test this emerging imaging method to 
determine its advantages and tradeoffs and compare them to the performance of
other existing techniques.  Preliminary results of observign with
a prototype gaussian pupil mask can be found in Debes et al. (2002).

In this proceeding we report further lab tests of the performance of GAPMs
compared to lyot coronagraphs with similar throughput as well as testing 
the new technique of combining a GAPM with a coronagraphic image plane mask.  
We theoretically compare different techniques with the same throughput on the 
Hubble Space Telescope to determine what may be useful in a real spacecraft.
  \begin{figure}
   \plottwo{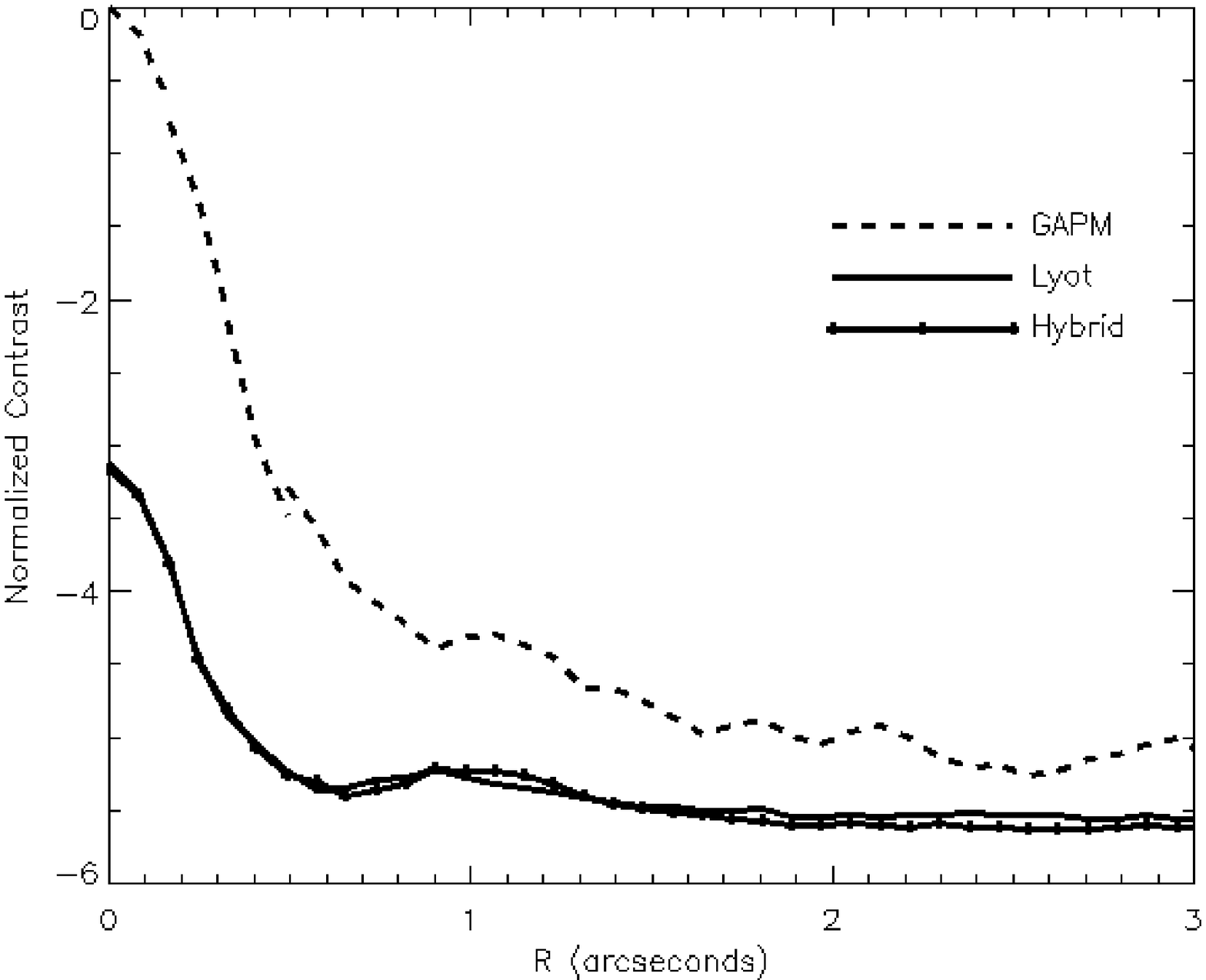}{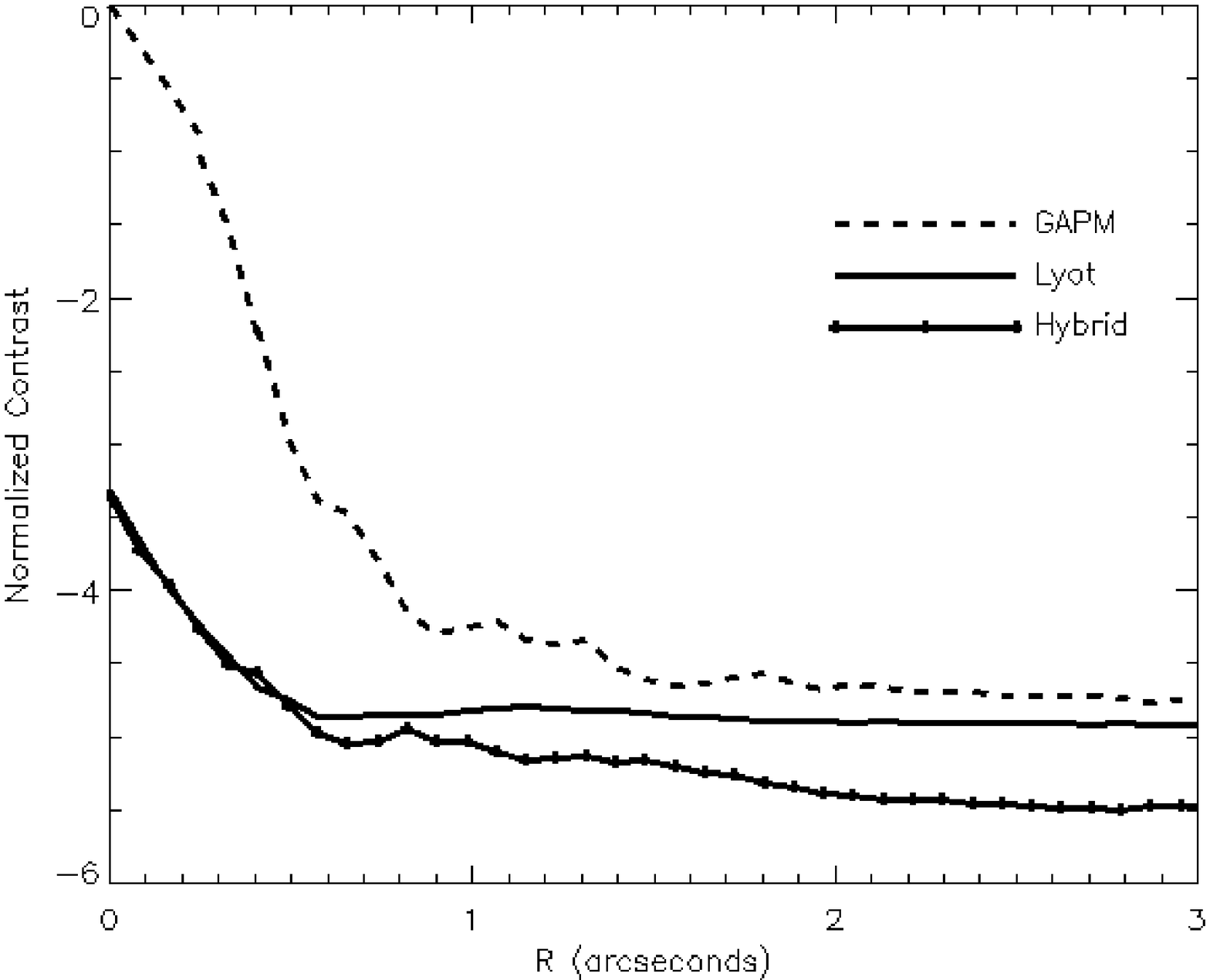}
   \caption{ \label{fig8} (left) Azimuthally averaged PSF profiles of the idealized GAPM,
hybrid, and Lyot coronagraph designs taken in the J band in the lab with PIRIS.
 (right) The same for the new designs for 
Mt. Wilson.  The best contrast achieved is $\sim$3$\times\ 10^{-6}$ at 10 
$\lambda/D$ or 1$^{\prime\prime}$.
}
   \end{figure}
\section{Equal Throughput Comparisons}
\label{equal}
A fairer comparison between a Lyot coronagraph and the GAPM is to have equal 
throughput designs and compare their contrast levels.  As part of  
lab experiments that we are performing we compared two new GAPM designs with 
a Lyot coronograph that had a comparable throughput.  Two types of designs 
were tested, idealized apertures with no secondary structure (20\% 
throughput) and realistic
masks that will be used for future observing (30\% throughput), which have
two gaussian aperture per quadrant, avoiding support structure.   
Unsaturated images GAPMs were taken to predict the flux
for the longer, saturated images.  We took exposures that had on the order of 
10$^{7}$
counts in order to image the PSF.  We found that this was insufficient to get 
high S/N on the fainter 
portions of the PSF and we estimate that beyond 1-2$^{\prime\prime}$ the read
noise begins to dominate.  Longer integrations are planned.  

For the 
coronagraphic modes we used a gaussian transmission focal plane mask with a
FWHM of 500$\mu m$ ($\sim 11\lambda/D$).
Short exposures were taken without the mask for estimating the peak flux of an
unblocked point source for a given exposure time.  The mask was then carefully
aligned to within 1 pixel to block the point source.  

Figure 1 shows the results of these lab tests.  In both cases the 
hybrid designs perform as well or better, both reaching 
$\sim2 \times\ 10^{-6}$.  The flattening of the profile suggests that the
observations were hitting the read noise limit of the PICNIC detector on PIRIS.
 Other experiments that were done
where the mask was misaligned by several pixels presented dramatically worse
results, underscoring the need for subpixel alignment and stability over an
observation.  This points to the utility of the GAPMs alone for quick surveys
and hybrid or Lyot designs for deeper searches.
   
\section{Testing other types of masks}

Given the large number of potential ideas for high contrast imaging, and due to
a lack of many lab tests of these designs, a way to model some of the various
contrast degradations present in a real space mission would be useful for 
determining what designs are better suited for future TPF type missions.
A strong test would be to compare the performance of 
the different designs on the HST, where many of these different errors are 
well modeled.  Tinytim, the PSF modeling software used by the Space Telescope
Institute, has accurate wavefront error maps of the telescope (Krist 1995).  Using these
wavefront error maps as input to our models allows us to compare all of the
different designs with an equal footing of throughput.

Figure 2 shows
a horizontal cut along the axis of highest contrast for all of the designs.
In this preliminary simulation the band limited mask performs slightly better
than the other designs for the same amount of throughput.  This
is because the Lyot stop for the band limited mask is undersized
from its optimal shape by about 10$\%$.  This would block
some of the residual light that leaks through the focal plane
mask.  These simulations 
ignore mask errors in both the focal and pupil planes.  Possibly with 
PSF subtraction techniques a factor of 10-100 deeper contrast could be 
achieved, allowing some bright extrasolar planets to be observed with HST.
Further simulations need to be done to better understand the feasibility of
such observations.

\section{Conclusions}
\label{concl}
We have performed several simulations, lab tests, and telescope observations 
with GAPMs and Lyot coronagraphs in order to better understand
the interplay between theory and the reality of observations.  GAPMs alone 
provide an improvement over a simple circular aperture for quick high contrast
imaging.  The combination of a GAPM and a coronagraphic mask further supresses 
and improves the performance of the GAPM alone.  The masks are very sensitive 
to an accurate reproduction of shape and thus
need accuracies that may be as restrictive as sub-micron precision.  This is 
possible with new nanofabrication techniques that have been perfected at the
Penn State Nanofabrication facility, where future masks may be produced. 
Precisely fabricating these masks can potentially improve performance to 
the ideal limit for a mask provided it is above the scattered light
limit of the telescope, bringing it in line with Lyot coronagraphs
 of comparable throughput.  Simulations of different techniques on the HST
provide an avenue to test new technologies for future NASA missions such
as TPF, and show what can be possible from space.

\acknowledgements

Several important discussions with D. Spergel, M. Kuchner,
W. Traub, C. Burrows, and R. Brown were crucial in
our understanding of gaussian apertures, band limited masks, and HST image
performance.

J.D acknowledges funding by a NASA GSRP fellowship under grant NGT5-119. This
work was supported by NASA with grants NAG5-10617, NAG5-11427 as well as the
Penn State Eberly College of Science.  J. G. also acknowledges funding through Ball Aerospace Co..

\begin{figure}
  \plotone{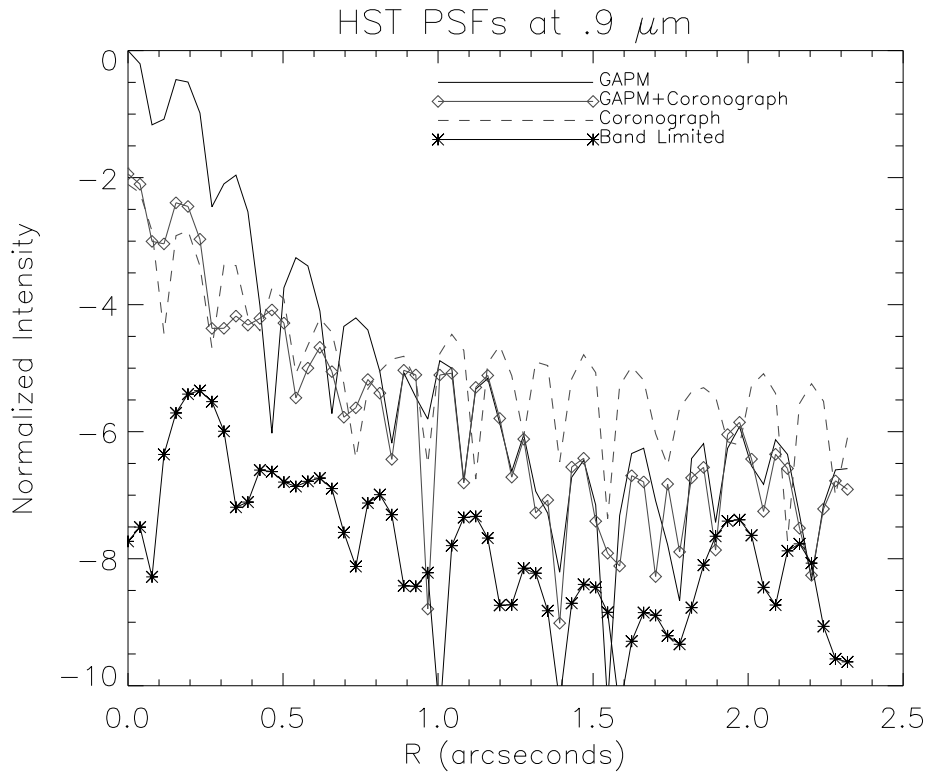}
   \caption{ \label{fig9} 
Horizontal cuts of the PSFs that would be generated on HST at .9$\mu m$
 for various designs postulated as useful for extrasolar planet detection. 
}
   \end{figure}

\end{document}